\documentstyle{article}

\def \bea {\begin{eqnarray}}
\def \eea {\end{eqnarray}}

\newcommand{\be}{\begin{equation}}
\newcommand{\ee}{\end{equation}}

\begin{document}    

\title{\Large \bf Symmetries of particle motion}

\author{Roy Maartens$^1$ and David Taylor$^2$\\
$^1${\small \it School of Mathematical Studies, Portsmouth University,
England}\\
$^2${\small \it Dept. Computational and Applied Mathematics,
University of the Witwatersrand, 
 South Africa}}

\date{}

\maketitle

\begin{abstract}

We define affine transport lifts on the tangent bundle by associating 
a transport rule for tangent vectors with a vector field on the base
manifold. The aim is to develop tools for the study of kinetic/
dynamical symmetries in relativistic particle motion. The transport 
lift unifies and generalises 
the various existing lifted vector fields, with clear geometric
interpretations. We find the affine dynamical symmetries of general
relativistic particle motion, and compare this to previous results 
and to the alternative concept of ``matter symmetry".

\end{abstract}

\section{Introduction}
\setcounter{equation}{0}

Vector fields on the tangent bundle $TM$, arising as the lifts of 
vectors or
of transformations on the base manifold $M$, have been defined and 
applied in
differential geometry, Lagrangian mechanics and relativity.  
For example the
complete (or natural or Lie), horizontal and vertical lifts
\cite{l1}--\cite{l4}, the projective
and conformal lifts \cite{l5} and the matter symmetries of 
Berezdivin and
Sachs \cite{l6}, \cite{l7}.
Our aim is to find a more general way of lifting from $M$ to $TM$ 
than the
usual definitions that involve only the vector field on $M$, and 
possibly
the connection on $M$.  In fact the matter symmetries of \cite{l6}
are a step in this direction.  We generalise this concept in a
way that gives a clear geometric foundation to all the lifts 
previously
defined, and to new lifts which can be defined.

The main idea \cite{l7a} 
is to associate a transport rule for tangent vectors 
with a
vector field on $M$.  This defines a vector field on $TM$ -- the 
transport
lift.   The class of {\em affine transport lifts} (ATL's)  
generalises all
previously defined lifts in a unified and geometrical way.  
We find conditions
under which ATL's are dynamical symmetries for particle 
trajectories in
(semi-) Riemannian manifolds.

\section{Local geometry of the tangent bundle}
\setcounter{equation}{0}

We give a brief summary of the relevant local differential
geometry of the tangent bundle assuming only a knowledge of 
basic tensor
analysis on manifolds.
Consider a (semi-) Riemannian $n$-manifold $(M,g)$ with local 
coordinates
$x^a$ and metric connection $\Gamma^a{}_{bc}$ (Christoffel symbols).
The tangent bundle $TM$ is the union of all tangent spaces (fibres)
$T_xM$, $x\in M$.
In relativistic kinetic theory (RKT) the phase space arises out of 
$TM$
by restriction to future-directed, 
non-spacelike tangent vectors \cite{l8}.

Local coordinates $x^a$ on $M$ induce local 
coordinates $\xi^I=(x^a,p^b)$
on $TM$, where $p^a$ are the coordinate components of the vector
$p=p^a\partial/\partial x^a$.
Any smooth vector field on $TM$ can be
expressed covariantly via the anholonomic ``connection basis"
$\{H_a,V_b\}$ of horizontal and vertical vector fields \cite{l4}:
\be
H_a={\partial\over\partial x^a}-\Gamma^b{}_{ca}p^c\;{\partial\over 
\partial
p^b}\,,~~~~ V_a={\partial\over \partial p^a}\,.
\ee
The Lie brackets of the basis vectors are
\bea
&&[V_a, V_b]=0 \,, \\
&&[H_a,V_b]=\Gamma^c{}_{ab} V_c \,, \\
&&[H_a,H_b]=-R^d{}_{cab} p^c V_d \,,
\eea
where $R_{abcd}$ is the Riemann curvature tensor.
The vector field
\be
\Gamma=p^aH_a
\ee
has integral curves on $TM$ which are the natural lifts of geodesics 
on $M$.
$\Gamma$ is called the geodesic spray or, in RKT, the Liouville vector
field.

For a vector field $Y=Y^a(x)\partial/\partial x^a$ 
on $M$, various lifted vector
fields have been defined on $TM$:
\bea
\mbox{Horizontal lift}&:&Y\to\overline Y=Y^a(x)H_a \,, \\
\mbox{Vertical lift}&:&Y\to\widehat Y=Y^a(x)V_a \,, \\
\mbox{Complete lift}&:&Y\to\widetilde Y=Y^a(x)H_a+
\nabla_bY^a(x)p^bV_a \,, \\
\mbox{Iwai's lift}&:&Y\to Y^\dagger=\widetilde Y-2\psi(x)p^aV_a \,,
\eea
where $\nabla$ is the covariant derivative,
and $\psi$ is proportional to $\nabla_aY^a$ in (2.9).

We can also define the vertical lift of a rank-2 tensor field
\cite{l1}
\be
A\to\widehat A=A^a{}_b(x)p^bV_a \,,
\ee
with a special case being the Euler vector field \cite{l4}
\be
\Delta=\widehat\delta=p^aV_a \,.
\ee

Matter symmetries in RKT have been defined \cite{l6}
in terms of a vector field $Y$ and a 
skew rank-2 tensor field $A$ on $M$:
\be
(Y,A)\to Y^a(x)H_a+A^a{}_b(x)p^bV_a \,,~~~~ A_{(ab)}=0\,,
\ee
where round brackets enclosing indices denote symmetrisation.

A dynamical system on $M$ is defined \cite{l2}, \cite{l3} 
by a congruence of
trajectories on $TM$.  The tangent vector field to these 
trajectories is
the dynamical vector field $\Gamma$, e.g. (2.5). A dynamical symmetry 
is a vector
field $\Sigma$ that maps trajectories into trajectories with possibly
rescaled tangent vector field.  Thus $(\exp\,
\varepsilon{\cal L}_\Sigma)\Gamma$ is parallel to $\Gamma$, 
where ${\cal L}$ is the Lie derivative. Hence
\be
{\cal L}_\Sigma\Gamma\equiv[\Sigma,\Gamma]=-\psi\Gamma   \,,
\ee
for some $\psi(x,p)$, is the condition for $\Sigma$ to be a dynamical 
symmetry.
The nature of the
rescaling depends on $\psi(x,p)$.
If $\psi=\psi(x)$, then the rescaling is constant on each fibre
$T_xM$.  If $\psi=0$, then there is 
no rescaling and $\Sigma$ is said to be
a Lie symmetry on $TM$.

\section{Transport lifts}
\setcounter{equation}{0}

Let $Y=d/d\sigma$ be a vector field on $M$ and $\Lambda$ a 
smooth local
rule governing the transport of tangent vectors along the integral 
curves
of $Y$.  Thus any $u^a$ at $x^a(\sigma)$ is mapped under $\Lambda$ to
$u^{\prime a}$ at $x^{\prime a}=x^a(\sigma+\varepsilon)$, with
$u^{\prime a}=
\Lambda^a(x,u;\varepsilon).$
This defines curves $(x^a(\sigma)\;,\; p^b(\sigma))$ in $TM$, with
\be
{dx^a\over d\sigma}=Y^a(x) \,,~~~~ {dp^a\over d\sigma}\equiv
\lambda^a(x,p)
={\partial\Lambda^a(x,p;0)\over
\partial\varepsilon} \,.
\ee
We can define a vector field on $TM$ with integral 
curves $(x^a(\sigma),
p^b(\sigma))$ given by (3.1).  
We call this the {\em transport lift\/} \cite{l7a} on $TM$ of
the vector field $Y$ and of the transport rule $\Lambda$ along $Y$.
The transport lift is given locally by
\bea
(Y,\Lambda) &\rightarrow & Y^a(x)\;{\partial\over\partial x^a}+
\lambda^a(x,p)\;
{\partial\over\partial p^a}\\
{}&=& Y^a(x)H_a+\left[\lambda^a(x,p)+
\Gamma^a{}_{bc}(x)p^bY^c(x)\right]
V_a \,.
\eea

The transport lift (3.2) combines the point transformations 
generated by
$Y$ on $M$
with the tangent vector transformations generated by $\Lambda$ on $M$.
In general, the transport rule $\Lambda$
along $Y$ is not defined purely by tensor fields on $M$.  However 
this 
is
the case for an affine transport rule, for which
$$
\Lambda^a(x,u;\varepsilon)=\Omega^a{}_b(x;\varepsilon)u^b+
K^a(x;\varepsilon).
$$
Thus the {\em affine transport lift} (ATL) of $(Y,\Lambda)$ on 
$M$ has the
form \cite{l7a}
\be
Y^{(A,k)}=Y^a(x)H_a+[A^a{}_b(x)p^b+k^a(x)]V_a\,,
\ee
where
\bea
&&A^a{}_b(x)=\omega^a{}_b(x)+\Gamma^a{}_{bc}(x)Y^c(x) \,, \\
&&\omega^a{}_b(x)={\partial
\Omega^a{}_b(x;0)\over
\partial\varepsilon}\,,~
k^a(x)={\partial K^a(x;0)\over\partial\varepsilon} \,.
\eea
It follows that $k$ is a vector field on $M$, whereas $\omega$ is not 
a tensor
field unless $Y=0$.  Furthermore, $A$ as defined by (3.5) is a 
tensor field,
and the vertical component in (3.4) therefore transforms covariantly.
The transport rule $\Lambda$ is thus covariantly determined by $A$
and $k$.

By (3.4), the integral curves of $Y^{(A,k)}$ satisfy
\bea
{dx^a\over d\sigma}&=&Y^a(x) \,,\\
{dp^a\over d\sigma}&=&\omega^a{}_b(x)p^b+k^a(x)
\nonumber\\
&=& [A^a{}_b(x)-\Gamma^a{}_{bc}(x)Y^c(x)]p^b+k^a(x) \,.
\eea
We can rewrite (3.8) as
$$
{Dp^a\over d\sigma}=A^a{}_bp^b+k^a \,,
$$
which shows that $A$ and $k$ determine the rate of change of 
tangent vectors
under $\Lambda$ relative to parallel transport.  In the case $k=0$, 
we get a
particularly simple interpretation of $A$:
\be
A^a{}_bu^b=\nabla_Yu^a ~~\hbox{or}~~ A(u)=\nabla_Yu \,,
\ee
for all $u$ along $Y$.   This equation is important for the geometric
construction of lifts (see below).
The class of {\em linear transport lifts} (LTL's) arises as the 
special 
case 
$k^a=0$, and we write
$$
Y^{(A)}\equiv Y^{(A,0)} \,.
$$
LTL's encompass all previously defined lifts apart from
the vertical lift (2.7).

Now from (3.4) we get
\be
\alpha Y^{(A,k)}+\beta Z^{(B,\ell)}=
(\alpha Y+\beta Z)^{(\alpha A+\beta B,
\alpha k+\beta\ell)} \,,
\ee
for any scalars $\alpha,\beta$ on $M$.  (Note that $A$ and 
$B$ depend,
respectively, on $Y$ and $Z$.  In particular, this means that in
general the taking of the affine transport lift is not a 
linear operation.)
Thus the ATL's form a linear subspace. Furthermore, (2.2--4) give
\be
[Y^{(A,k)},Z^{(B,\ell)}]=[Y,Z]^{(C,m)}\,,
\ee
where
\bea
&&C=\nabla_YB-\nabla_ZA-[A,B]-R(Y,Z) \,,\\
&&m=\nabla_Y\ell-\nabla_Zk-A(\ell)+B(k) \,.
\eea
$C$ is a rank-2 tensor field on $M$, with $[A,B]$ the tensor 
commutator, and 
$R(Y,Z)^a{}_{b}=R^a{}_{bcd}Y^cZ^d$.
By (3.10--13), the ATL's form a Lie algebra.   The LTL's
are a subalgebra (but not an ideal).

Before limiting ourselves to the linear case, we regain the vertical 
lift (2.7)
of a vector field:
\be
\widehat Z=0^{(0,Z)} \,.
\ee
By (3.4), if $Y=0$, we regain the vertical lift (2.10) of the rank-2 
tensor
field $A$:
\be
0^{(A)}=A^a{}_bp^bV_a=\widehat A \,.
\ee
$0^{(A)}$ generates a $GL(n)$ transformation on each fibre:
$p^a\rightarrow p^{\prime a}=(\exp\epsilon A)^a{}_bp^b.$
Thus on each fibre $T_xM$, $A^a{}_b(x)$ is an element of the Lie 
algebra
$g\ell (n)$.   By restricting $A^a{}_b(x)$ to a particular Lie 
subalgebra, we 
see
that $0^{(A)}$ generates gauge transformations of the 
corresponding Lie
group.

In order to regain the horizontal lift (2.6) of a vector field, we 
require
that the transport rule $\Lambda$ be {\it parallel transport} along
$Y$ for all $u$. 
When the transport rule $\Lambda$ is chosen to be {\it Lie transport}
(``dragging along"), we regain the complete lift (2.8):
\be
\overline Y=Y^{(0)}\,,~~~\widetilde Y=Y^{(\nabla Y)} \,.
\ee
Thus we are able to regain in a unified and geometric way, 
the standard lifts
of vectors and rank-2 tensors via the concept of ATL's.  Using the 
general
Lie bracket relation (3.11), we can easily regain the known 
Lie brackets
\cite{l1}--\cite{l4}
amongst the three standard vector lifts:
\bea
&&[\overline Y,\overline Z]=[\overline{Y,Z}]-\widehat{R(Y,Z)}\,,~~~~
[\overline Y,\widehat Z]=\widehat{\nabla_YZ} \,, \\
&&[\overline Y,\widetilde Z]=[\overline {Y,Z}]+\widehat{S(Y,Z)}\,,~~~~
[\widehat Y,\widehat Z]=0 \,, \\
&&[\widehat Y,\widetilde Z]=[\widehat{Y,Z}]\,,~~~~
[\widetilde Y,\widetilde Z]=[\widetilde{Y,Z}] \,,
\eea
where $S(Y,Z)^a{}_b =({\cal L}_Z
\Gamma^a{}_{cb})Y^c$. Note that the sets of 
vertical and
complete lifts each form a Lie algebra, but the horizontal lifts 
do not on a
curved manifold.  
By (3.11), the vertical lifts form an ideal in the algebra
of ATL's, but the complete lifts do not.

We now show \cite{l7} that the LTL's also include the 
matter symmetry vector fields of
RKT. Berezdivin and Sachs define a matter symmetry as a 
vector field on $TM$
that leaves the distribution function $f$ unchanged.  
This vector field
connects points in $TM$ where the distribution of matter is the same.
Geometrically, this implies that an observer at $x$ with local 
Lorentz frame 
$F$
will measure $f$ on the tangent fibre 
$T_xM$ to be the same as an observer
at $x'$ with Lorentz frame $F'$ measuring $f'$ on $T_{x'}M$.   Thus
matter symmetries arise in the class of 
LTL's out of the requirement that the
transport rule $\Lambda$ be {\it Lorentz transport} along $Y$. Hence
any vector
transforms according to a representation of the 
Lorentz group $SO(1,3)$ along 
$Y$.
Given an orthonormal tetrad $\{E_a\}$, 
we have $E_a\cdot E_b=\eta_{ab}\equiv\;
\mbox{diag}\;(-1,1,1,1).$
Now the tetrad components of any vector transform as
$u^{\prime a}=\Lambda^a(u,x;\varepsilon)=
\Omega^a{}_b(x;\varepsilon)u^b$
where $\Omega\in SO(1,3)$.  Thus  $\Omega$ preserves $\eta$.
Differentiating and noting that $\Omega^a{}_b(x;0)=\delta^a{}_b\,$,
we get
\be
\omega_{(ab)}=0~\Rightarrow~ A_{(ab)}=0 \,,
\ee
where $\omega$ is defined by (3.6). 
This is the condition in (2.12) for $Y^{(A)}$ to
be a matter symmetry -- or ``Lorentz lift".
The matter symmetries form a Lie algebra, since by (3.12),
$C$ is skew if $A$ and $B$ are.

Iwai's lift (2.9) arises as the LTL which is the 
lift of {\it conformal Lie
transport}.  However Iwai defines his lift for $Y$ 
a projective collineation
or conformal Killing vector, whereas the class of ATL's 
generalises this
to any $Y$:
\be
Y^\dagger =Y^{(\nabla Y-2\psi\delta)} \,.
\ee
The generalised Iwai lifts form a Lie algebra:
$$
[Y^\dagger,Z^\dagger]=[Y,Z]^\dagger ~~
\mbox{where} ~~ \psi_{[Y,Z]}=
{\cal L}_Y\psi_Z-{\cal L}_Z\psi_Y \,.
$$
This generalises Iwai's result \cite{l5} to the case of 
arbitrary $Y,Z$.

\section{Dynamical and matter symmetries}
\setcounter{equation}{0}

In searching for a dynamical 
symmetry $\Sigma$ obeying the condition (2.13) with
$\Gamma$ the geodesic spray (2.5),
it is usually assumed that $\Sigma$ arises purely 
from a vector field on the
base manifold $M$ -- for example, $\Sigma=\widetilde Y$ or 
$Y^\dagger$.
Transport lifts 
open up the possibility of generalising dynamical symmetries
to the case where not only a vector field, but also a transport 
law for
tangent vectors, is used to generate transformations of the dynamical
trajectories.  In the case of affine transport laws, 
this means looking at the
ATL's.
Unfortunately, as we shall show, the
dynamical symmetry condition reduces the ATL to a vector lift -- 
in fact to
$Y^\dagger$ \cite{l7a}.  
At least this gives a foundation to the ad hoc ansatz
of Iwai.

We examine now the conditions under which an ATL is a dynamical
symmetry.  By (2.13)
this gives
\be
[Y^{(A,k)},\Gamma]=-\psi\Gamma \,,
\ee
where $\Gamma$ is given by (2.5). Then (4.1) implies $k^a=0$, 
which is the
restriction to the class of LTL's. A further implication of (4.1) 
is that
\be
A_{ab}=\nabla_bY_{a}-\psi g_{ab} \,.
\ee
From (4.2) it is clear that $\psi$ is restricted to $\psi=\psi(x)$, 
and
\be
{\cal L}_Y\Gamma^a{}_{bc}\equiv \nabla_c\nabla_bY^a-R^a{}_{bcd}Y^d=
\delta^a{}_{(b}\nabla_{c)}\psi \,.
\ee
By (4.3), $Y$ is a projective collineation vector \cite{l2}, 
\cite{l3},
and together with (4.2) this means that the ATL is reduced to Iwai's
projective lift (2.9):
$$
[Y^{(A,k)},\Gamma]=-\psi\Gamma ~~\Rightarrow~~ Y^{(A,k)}=
Y^{(\nabla 
Y-\psi\delta,0)}=
Y^\dagger \,.
$$
Thus we see that any affinely based dynamical symmetry arises from a
projective collineation vector.  Furthermore,
the ansatz introduced by Iwai in fact arises as the condition
for an ATL to be a dynamical symmetry.  
Any attempt to generalise Iwai's ansatz
would require a fully nonlinear transport rule $\Lambda$.

Matter symmetries provide a different, and more physically based,
approach to symmetries of particle motion, but are correspondingly 
more difficult to analyse. By (2.12), we find that \cite{l7}
\be
[Y^{(A)},\Gamma]=\left(A^a{}_b-\nabla_bY^a\right)p^b\,H_a
+\left(R^a{}_{bcd}Y^d-\nabla_cA^a{}_b\right)p^bp^c\,V_a \,.
\ee
Thus a matter symmetry is more general than a dynamical symmetry,
and reduces to the latter only if
\be
A_{ab}=\nabla_{[b}Y_{a]}\,,~~\nabla_{(b}Y_{a)}=\psi g_{ab}\,,~~
{\cal L}_Y\Gamma^a{}_{bc}=0 \,.
\ee
These conditions imply that $Y$ is a conformal Killing and
affine collineation vector, i.e. a homothetic vector ($\psi$ is
constant), and $A$ is the bivector $dY$.

\section {Conclusion}

By generalising the concept of lifting point transformations to 
include
tangent vector transport, we have defined the class of ATL's on the 
tangent
bundle.  The ATL's include all previous lifts, thus unifying many 
results
into a single framework, with clear geometric interpretations.  The
generalisation introduced by the ATL concept includes in 
particular the
matter symmetries of RKT, and the lifts introduced ad hoc
by Iwai.   The projective lift of Iwai is shown to be the unique ATL
which is a dynamical symmetry on (semi-) Riemannian manifolds.  The 
matter symmetries provide a very different concept of invariance --
see \cite{l7} for a full discussion. They
coincide with
dynamical symmetries only in the special case that $Y$ is 
homothetic and $A=dY$.

Applications of the ATL formalism beyond RKT are possible.
It may also be useful in the study of symmetries in gauge 
field theories,
since $Y^{(A)}$ generates gauge transformations along $Y$ if $A$ 
is in the 
gauge
Lie algebra at each point.
The formalism could also be generalised to other fibre bundles.  
For example,
an ATL on the $({r\atop s})$ tensor bundle $T^r_sM$ arises 
when $\Lambda$
transforms $({r\atop s})$ tensors along $Y$.  With modifications, 
the formalism
would also carry through to the tangent bundle of a manifold 
with torsion.

\end{document}